\documentclass[ijgi,article,accept]{Definitions/mdpi}

\usepackage{tikz,xcolor}

\definecolor{lime}{HTML}{A6CE39}
\DeclareRobustCommand{\orcidicon}{%
	\begin{tikzpicture}
	\draw[lime, fill=lime] (0,0) 
	circle [radius=0.16] 
	node[white] {{\fontfamily{qag}\selectfont \tiny ID}};
	\draw[white, fill=white] (-0.0625,0.095) 
	circle [radius=0.007];
	\end{tikzpicture}
	\hspace{-2mm}
}

\foreach \x in {A, ..., Z}{%
	\expandafter\xdef\csname orcid\x\endcsname{\noexpand\href{https://orcid.org/\csname orcidauthor\x\endcsname}{\noexpand\orcidicon}}
}

\firstpage{1} 
\pubvolume{xx}
\issuenum{1}
\articlenumber{5}
\pubyear{2020}
\copyrightyear{2020}
\history{Received: date; Accepted: date; Published: date}

\Title{AIRSENSE-TO-ACT: A Concept Paper for COVID-19 Countermeasures based on Artificial Intelligence algorithms and multi-sources Data Processing}

\Author{ Alessandro Sebastianelli $^{1,}$\orcidB{}, Francesco Mauro$^{1,}$, Gianluca Di Cosmo$^{1,}$, Fabrizio Passarini $^{2}$\orcidD{}, Marco Carminati $^{3}$\orcidC{}, Silvia Liberata Ullo $^{1}$\orcidA{}}

\AuthorNames{Alessandro Sebastianelli, Francesco Mauro, Gianluca Di Cosmo, Silvia Liberata Ullo,  Fabrizio Passarini, Marco Carminati}

\address{%
$^{1}$ \quad Engineering Department, University of Sannio, Benevento, Italy; ullo@unisannio.it; sebastianelli@unisannio.it\\

$^{2}$ \quad Department of Industrial Chemistry Toso Montanari, University of Bologna, Bologna, Italy; fabrizio.passarini@unibo.it\\

$^{3}$ \quad Dipartimento di Elettronica, Informazione e Bioingegneria, Politecnico di Milano, Milan, Italy;
marco1.carminati@polimi.it\\
}

\corres{Correspondence: ullo@unisannio.it}

\abstract{Aim  of this paper is the description of a new tool to support institutions in the implementation of targeted countermeasures, based on quantitative and multi-scale elements, for the fight and prevention of emergencies, such as the current Covid-19 pandemic. The tool  is a centralized system (web application),  
single multi-user platform, which relies on  Artificial Intelligence (AI)  algorithms for the processing of heterogeneous data, and which can produce an output level of risk. The model includes a specific neural network which will be first trained to learn the correlation between  selected inputs, related to the case of interest: environmental variables (chemical-physical, such as meteorological), human activity (such as traffic and crowding), level of pollution (in particular the concentration of particulate matter), and epidemiological variables related to the evolution of the contagion. The tool realized in the first phase of the project will serve later both as a decision support system (DSS) with predictive capacity, when fed by the actual measured data, and as a simulation bench performing the tuning of certain input values, to identify which of them lead to a decrease in the degree of risk. In this way, the authors aim to design different scenarios to compare different restrictive strategies and the actual expected benefits, to adopt measures sized to the actual need, and adapted to the specific areas of analysis, useful to safeguard human health, but also the economic and social impact of the choices.
}


\begin{document}
\section{Introduction}
In this concept paper authors aim to focus on the challenges posed by the Covid-19 pandemic and present a multidisciplinary and quantitative approach to respond to the sanitary and economical crisis. The proposed model is named AIRSENSE-TO-ACT, and is the description of a project submitted  to the 2020 FISR Call of MIUR (Italian Ministry of University and Research), issued to collect solutions, related to the diffusion of the Covid-19 pandemic, able to  contain its effects and offering a novel way for the management of the reorganization of activities and processes.

The model is based on the fusion of heterogeneous data coming from different sensors: on board of satellites and/or  positioned on ground platforms, both mobile and fixed, in addition with other public data extracted from databases, all jointly processed  through the application of Machine Learning (ML) algorithms, and  the employment and comparison of macro- and micro-systems of analysis.

Some data can be indeed extracted from public databases (i.e. epidemiological information, number of infected, etc.), or retrieved from satellite images freely downloadable, such as those of the European Space Agency (ESA) Copernicus mission, since public databases  have further increased in number and type of data, following the COVID-19 emergency. 

Yet, to improve the model and the analyses carried out, it is foreseen the possibility to carry out data collection campaigns through ground-based networked sensors for instance, to validate the developed model, in particular in the Italian areas where the emergency has shown more critical issues, such as in the Po Valley, and specifically in the Emilia-Romagna and Lombardy regions. Local and punctual measures with better spatial and temporal resolution will increase the effectiveness of the fight against the spread of the contagion, allowing to focus the analysis from macro-areas covered by satellite,  to micro-areas in order to  increase the "granularity" of the data and to reduce the reaction time on the decisions to be taken. 

It is necessary to underline that a crucial aspect will concern the creation of the datasets with which the  artificial network underlying the DSS needs to  be trained, since based on previous experience, this activity may take up to 3/4 months.

However, it is also worth pointing out that,  although developed to combat the COVID-19 pandemic in Italy, this model can be applied to other emergency situations, where air quality has health implications, above all in other countries with similar economic-infrastructural characteristics.

The proposal falls mainly within the scope of risk prevention, developing solutions to counteract and contain the effects of COVID-19 and any future pandemics. However, thanks to the versatility of the proposed tool, it is also of immediate use for the response to other emergencies, to develop tailored solutions and for the management of the organization of activities and processes, relating to the phase of overcoming the phenomenon in safety conditions.

The idea was born because the actual measures of lockdown taken by Governments and local Institutions (Regions and Municipalities) are always \textit{a-posteriori} decisions, where increasing levels of lockdown are activated, based on the number of infected, hospitalized and dead people,  up to a generalized lockdown, like the one imposed in Italy from the beginning of March until almost the end of June, and which seems to come in the next weeks of October again in Italy. These measures imitated what was decided in Wuhan area, in China, to contain COVID-19, having proved the positive impact of lockdown on the number of infections \cite{Lau2020}, \cite{Wang2020}.

Yet, to protect human health, but at the same time safeguard socio-economic aspects, these methods of intervention are not appropriate  for their negative implications \cite{Nicola2020}. The type of interventions must learn from the past to take \textit{a priori} decisions, which have a minimal impact on commercial activities but are effective in limiting the spread of the virus. 
An interesting analysis of responses to COVID, carried out in Indonesia during the first months of the 2020, is presented in \cite{Djalante2020}, where five recommendations are given to face efficiently the pandemic situation. Micro actions are highlighted and interventions step-by-step analysed. The only limit is related to the methodology, based on traditional tools.

Artificial Intelligence algorithms instead can lend a hand in this sense, managing to capture the hidden interactions between data, and providing the possibility of their corresponding use for a micro- and macro-analysis of the phenomenon, allowing localized and targeted interventions.

As already mentioned above, the proposed model aims to combine satellite data, and data acquired through ground platforms, both mobile and fixed, related to the concentration of some pollutants such as NOx, PM10, PM2.5, meteorological data, air mass displacement, chemical-physical parameters such as temperature and humidity, with other reference data such as population mobility, epidemiological data (number of infected), number of places still available in intensive care (global values or per-hospitalization points, distributed throughout the regions, nationally, etc.), the concentration of residents per $km^2$, the degree of implemented  lockdown, and so on.
Several studies analyze the impact and correlation of the mentioned factors on COVID-19
\cite{Nicola2020, Djalante2020,  Tosepu2020, Lau2020a, Bashir2020,  Kraemer2020, Bashir2020a, Kim2020}. 

The idea is to create a Decision Support System (DSS) able to produce as output the level of risk, based on the correlation between the selected data. The choice to base the proposed model on Artificial Intelligence algorithms stems from the enormous amount of data which must be taken into account, and the need to analyze the correlations, sometimes hidden, between this information. In fact, completely different causality structures emerge from the literature when the various factors are considered in combination. For example, the rise in temperature seems to lead to a reduction in the diffusion of COVID-19, but it depends also on the humidity values, since it has been found that high temperature values with high humidity values do not stop the spread of COVID, but on the contrary facilitate it.
In \cite{aircovid} the Covid Risk Weather (CRW) parameter is introduced,  an index to evaluate the relative COVID-19 risk due both to weather and air pollution. The CRW can be used to compare the relative changes in reproductive number for the disease due to the weather factors (average and diurnal temperature, ultraviolet (UV) index, humidity, pressure, precipitation) and air pollutants (SO2 and Ozone). 
The authors highlighted  in   \cite{aircovid} that  warmer temperature and moderate outdoor UV exposure may offer a modest reduction in reproductive number. However, UV exposure can not fully contain the transmission of COVID-19. If on one hand both high temperature and solar radiation are able to speed up the inactivation rate, on the other high relative humidity may promote the diffusion rate \cite{aircovid, aereosol}.
Still from the literature, in the diffusion of respiratory diseases, the interaction between multiple elements: pollution, high population density, overcrowding, is analyzed and recent studies have shown a correlation between high concentrations of fine particulate matter and  COVID-19 diffusion  \cite{Li2020, Jordan2020}.  However, the subject is still much debated, and further multidisciplinary investigations are in progress. Some additional  considerations will be given ahead in this work.\\

Based on all the above considerations, the proposed approach aims to  develop a novel model for the cooperative fusion of extremely heterogeneous data, both in terms of nature and source (epidemiological data, environmental data, and data related to the human activities), and in terms of spatial (km to m) and temporal (days to seconds) sampling, to capture the  extremely complex dynamics, difficult to identify with other traditional methodological tools.  As highlighted before, the algorithms of AI are able to extract the features related to the hidden correlation between several elements, such as for instance the concentrations of atmospheric particulate matter, the meteorological trends and the virus spreading, to estimate the level of risk.\\

\noindent
The designed model is expected to operate on two levels of analysis:
\begin{itemize}
    \item \textbf{Macro analysis}: mainly through satellites (not limited to)  with the data collection   on wide  areas
    \item  \textbf{Micro analysis}:  through the use of fixed or dynamic networks for local-focus data collection
\end{itemize}
and initially it has been developed with particular attention to the most critical areas of Italy.

To the best of our knowledge, this proposal represents a significant progress compared to the state of the art. The only studies in progress, related to similar issues, concern the creation of the ESA RACE dashboard \cite{esarace},  published on the 5th of June 2020, which allows the use of satellite data to support the monitoring of commercial and productive activities, and on the other hand, the development of national and international projects such as PULVIRUS \cite{ispra1}, EpiCovAir and RESCOP \cite{rescop}, on the study of the correlation between suspended atmospheric particulate matter concentration and COVID-19. Our proposal integrates the above, and goes beyond their aims, by proposing a Decision Support System capable of producing the level of risk, but also useful to be used for simulations, for tuning the input values to establish which inputs to vary and how in order to obtain a lower level of risk.

Based on recent events and the studies of the researchers working on COVID-19 diffusion after that the contagion took place, it came out that there is a time lag  between the contagion moment and the manifestation of COVID-19 symptoms in the infected person (and its positive test results), a period evaluated between 14 and 21 days. Similarly, the closure of road traffic and the lockdown restrictions may have an effect on the reduction of pollution, but this happens  only after some days, and the elapsed time is different in different areas. Situation is worse  particularly in some parts of Italy, such as the Po Valley, which suffers from weather and orographic conditions that make it very problematic to shuffle air masses.  This latter point  has intrigued us, and some simulations and data analysis have been carried out to identify the number of days which best represents the elapsed time in this case for specific areas and critical regions. 
The authors started working on two case studies, that will be presented ahead in this work, where trends of NO2 have been analyzed in the Lombardy region (Italy) and Hubei region (China), where Wuhan is located, after the activation of the  lockdown measures.  Sentinel-5P data have been used to measure some pollutants and plot the NO2 values over time, and some analysis on the corresponding COVID-19 daily evolution has also been carried out, with considerations on the correlation between NO2 and COVID-19 evolution.

To take into consideration the presence of a delay in the cause-effect phenomena, the  type of the neural network necessary for our model should be able  to take into account  time series  deferred in time. Therefore, a particular network architecture called Long Short-Term Memory (LSTM) has been chosen for the creation of the DSS, as previously described. 
The final goal is  using historical data series to predict not only the level of risk at time T, but also in other subsequent instants, so that after being trained the network can also be used  as a transfer function to calculate the output (the expected risk level) by using new input data.

The complementarity of the three research groups, which will cover satellite monitoring and the development of data fusion and crowd monitoring models (Univesity of Sannio), miniaturization and networking of high-resolution sensors also on drones (Politecnico di Milano) and biochemical, health and environmental issues (Univesity of Bologna), demonstrates the strong interdisciplinary nature of the project and represents a guarantee of success for the creation, and subsequent validation, of the proposed system.  
\section{The Involved Factors in the DSS Design}

\begin{figure}[!ht]
    \centering
    \includegraphics[scale=0.5]{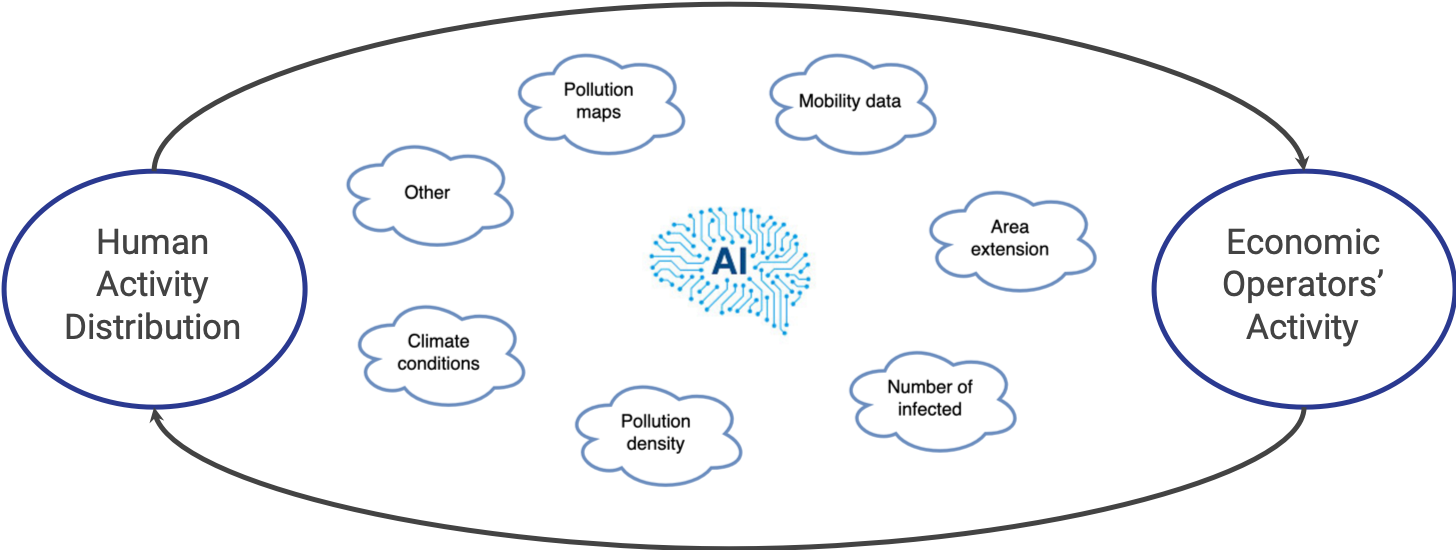}
    \caption{Data sources}
    \label{fig:data_sources}
\end{figure}

\noindent
Each country and its Government has a Department of Civil Protection managing the situations in case hazardous events happen. For instance, in Italy  the Department of Civil Protection \cite{protezcivile} activates specific interventions when several risks, such as seismic, volcanic, health, environmental, fire, and other, occur.
In the specific case of COVID-19 pandemic, the problem and its solutions have involved three main acting factors:

\begin{itemize}
    \item  \textbf{Pollution$\&$Population}: pollution, overpopulation, anomalies in climate conditions can lead to the disease
    \item \textbf{Spread of diseases}: the spread of diseases involves issues related to population and some geo-physical changes, but it can be mitigated by taking correct decisions
    \item  \textbf{Decisions taken}: decisions are taken by humans based on previous experience, but to take correct decisions the right information and  robust decision support systems are needed
\end{itemize}
Usually decisions are taken by humans, based on past experience and selected data, but to take a correct decision an objective and scientific method is needed. We have learned that under Covid pandemic, behaviors and/or needs such as gatherings, mobility, sharing of work spaces, etc. can heavily affect the number of infected people, and all the countries to reduce the pandemic have activated \textit{a posteriori} interventions, based on the number of infected and dead, through several attempts of consecutive lockdowns, each time more restrictive, but not always effective, or in any case with dramatic consequences on the social and economic framework. 
\\
Hence the idea of making a tool able to support Decision Makers  when multiple parameters are involved, realizing an  Artificial Intelligence (AI) Decision Support System (DSS), receiving information from multiple sources, and able to produce as output the degree of risk. The proposed system aims to provide an \textit{a priori} tool to take the right decisions, and the general scheme is represented in the  Fig. \ref{fig:data_sources}, where human activity contribution and economic operators' activity are highlighted, bot interacting under factors such as climate conditions, pollution density, mobility data, etc.,  resulting in the  number of infected and other effects, whose mutual interactions can be captured through  AI-based paradigms.

Clearly, such a system to be effective  must relies on the availability of data almost on real time, or with a low revisit time (i.e. Sentinel-5P and others), in order to help in realizing a continuous monitoring system, able to produce the right alert to manage jointly multiple risks such as in the case discussed, sanitary, environmental, and economic.

\section{Proposed architecture}
The proposed paradigm  is based on a  Long Short-Term Memory (LSTM) network, which allows the model to learn the temporal features from the training data. This type of network was invented to solve the problems of vanishing and explosion gradient of which the Recurrent Neural Networks (RNNs), a precursor of LSTMs, suffer \cite{LSTM_medium}.

\begin{figure}[!ht]
    \centering
    \includegraphics[scale=0.38]{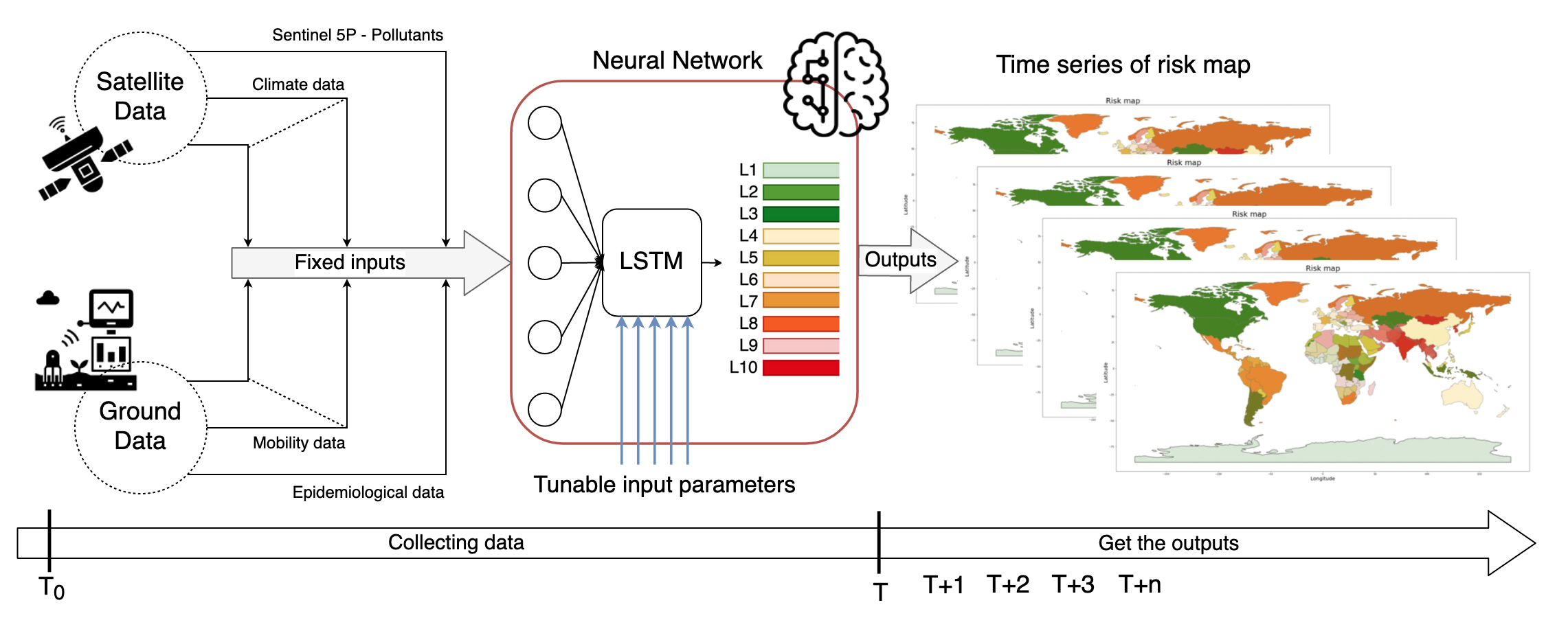}
    \caption{Block diagram of the adopted solution}
    \label{fig:block_diagram}
\end{figure}
\begin{figure}[!ht]
    \centering
    \includegraphics[scale=.25]{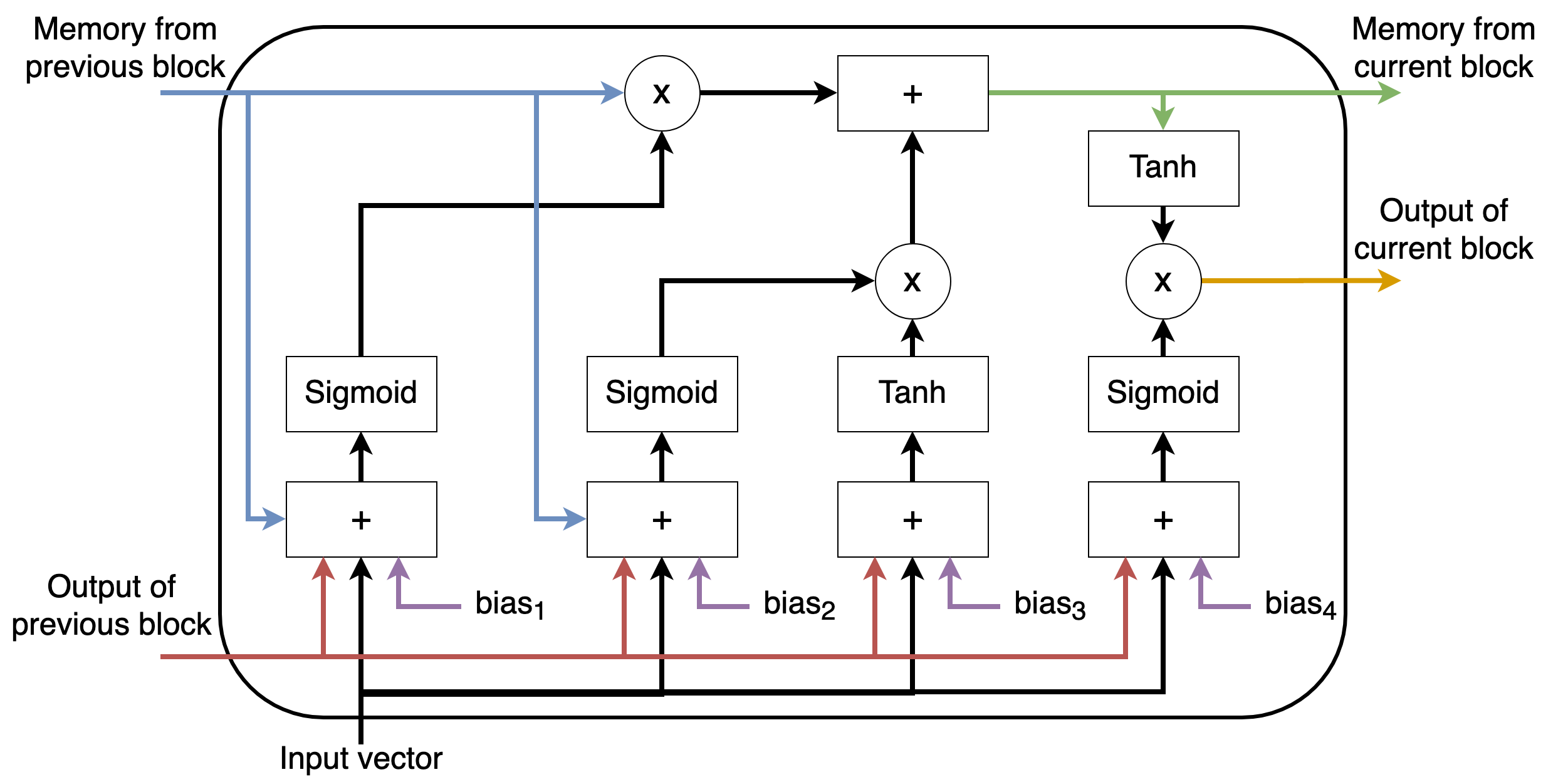}
    \caption{Elementary building block of the LSTM network}
    \label{fig:lstm_bb}
\end{figure}

The general block diagram of the adopted solution is presented in the Fig. \ref{fig:block_diagram}, while the elementary building block of the LSTM network is shown in the Fig. \ref{fig:lstm_bb}.

As it can be seen from the Fig.\ref{fig:lstm_bb}, each elementary building block of the LSTM network receives three inputs:  the input at the current time step, the output from the previous unit and the “memory” of the previous unit. Each single building block of the LSTM network makes a decision knowing the current input, previous output and previous memory, and it generates a new output and updates its memory.


The functioning of the proposed model can be splitted, as commonly it happens for neural networks, into two main phases: the training and the prediction. During the training phase the model is fed with input data and output data, both contained into the training set. That means that the model will be trained with a \textit{Supervised Learning} approach. During the supervised learning, the model needs both input and output data. Defined $X \in R^n$ a vector of inputs and $Y \in R^m$ a vector of outputs, then the model has to be  trained in order to find a correlation between X and Y, as shown by the equation \ref{eqn:supervised}.

\begin{equation}
    F \rightarrow Y = F(X)
    \label{eqn:supervised}
\end{equation}

For the model under analysis, we suppose that the input is a matrix $X \in R^{n, m}$, where $n$ represents the number of the different data sources and $m$ represents the size of the acquisitions in the time dimension. The same considerations can be done on the output, indeed $Y \in R^{p, q}$ where $p$ represents the dimensionality of the output in a fixed instant of time and $q$ represents the time dimension of the output. In other words, both the input and the output are matrices, the inputs will contain sensed data for a fixed time interval and the output will contains the prediction of the model for a fixed time interval.

The training process can be done with the well known Back Propagation and Gradient Descent techniques \cite{hecht1992theory, lecun1990handwritten, ruder2016overview}. In few words, for each training input, the model is updated by using the error and its derivative. The error is calculated trough an \textit{a priori}  defined Loss Function. The model will be trained in order to minimize this error. From this last statement, it can be better understood why both inputs and outputs are required in the training phase. For example, given a model $F$ and a set of data (X, Y) with $X \in R^{n, m}$ and $Y \in R^{p, q}$, during the training, for simplicity, a sample of the dataset is selected. Defined $X_i$ and $Y_i$ the sample, the training phase can be roughly represented by the equations: 

\begin{equation}
    y_{pred} = F(X_i)
    \label{eqn:mod1}
\end{equation}

\begin{equation}
    e = loss\_ function(y_{pred}, Y_i)
    \label{eqn:error}
\end{equation}

\begin{equation}
    \dot{e} = \partial_X loss\_ function(y_{pred}, Y_i)
    \label{eqn:der}
\end{equation}

Using the error and the the derivative of the error, respectively given by equations \ref{eqn:error}  and  \ref{eqn:der},  the model weights are updated, until reaching the minimum. After the training phase, the model should be able to generalize, and given new inputs it can generate the new outputs, using the internal state $F$.

\begin{figure}[!ht]
    \centering
    \includegraphics[scale=1]{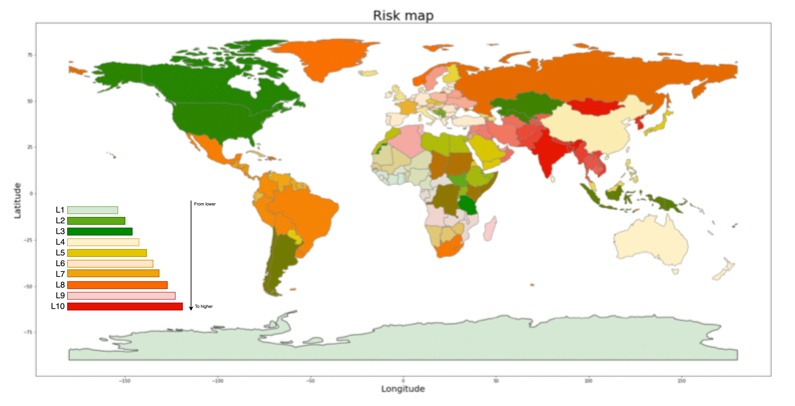}
    \caption{Example of risk map at low resolution}
    \label{fig:low_res_riskmap}
\end{figure}

The final output of the model will be a time series of risk maps, as shown in the Fig. \ref{fig:block_diagram}, with a resolution that can vary from low resolution intended as risk for a country to high resolution intended as risk for a city, or a smaller area. An example of a single risk map at low resolution is proposed in Fig. \ref{fig:low_res_riskmap}.

\section{State of the work}
As previously discussed, our proposal was born in response to a MIUR call on COVID and aims to create a multiscale analysis system based on AI algorithms, to provide \textit{a priori}  tool to decision-makers. This system is based on the combination of multi-source data with measurements obtained from fixed and mobile network sensors, from satellite information and proximity surveys, if necessary. Such a proposal required the integration of multi-disciplinary skills ranging from the development of sensor networks and satellite data processing to particle detectors, biochemical analysis of particulate matter, innovative strategies in the environmental field, and sophisticated computational technologies for treatment, analysis, and interpretation of data using Big Data technologies.
This paper intends to be a "Concept" paper, and much of the work must still be done, at least in reference to the realization of the neural network and to the creation of all necessary dataset.
Yet, some progress has been already done, and in the next sections the idea is to present the state of the work and give useful considerations and reasoning behind it.

\section{Measuring pollutants, air conditions and pandemic data}
Since the core of the idea involves the use of pollutants measurements and the infection data, beyond the involvement of other information, as already highlighted, the authors first selected a list of possible pollutants,  such as nitrogen dioxide (NO2), sulphur dioxide (SO2), formaldehyde (HCHO), ozone (O3), which can be retrieved through the use of Sentinel-5P, and other data (for instance, PM2.5, PM10, etc.) which can be retrieved by using other sources, at different multi-scale levels.

\begin{figure}[!ht]
  \centering
  \includegraphics[width=1\linewidth]{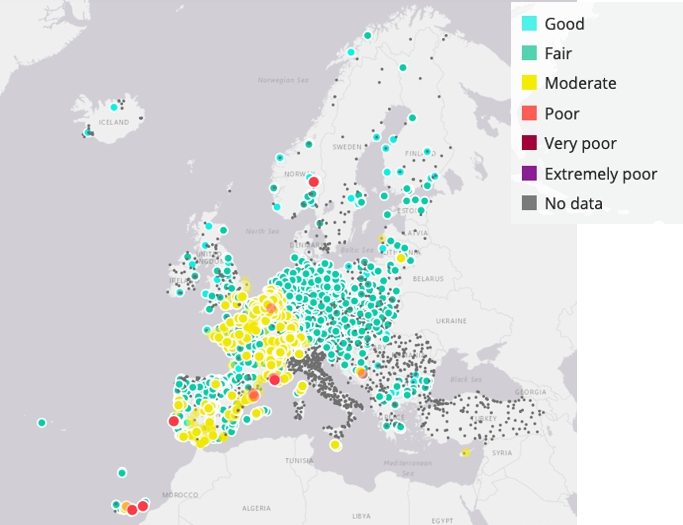}  
  \caption{Air quality map retrieved from the\\ European Environment Agency website \cite{europeanmap}}
  \label{fig:air_quality}
\end{figure}

\begin{figure}[!ht]
  \centering
  \includegraphics[width=1\linewidth]{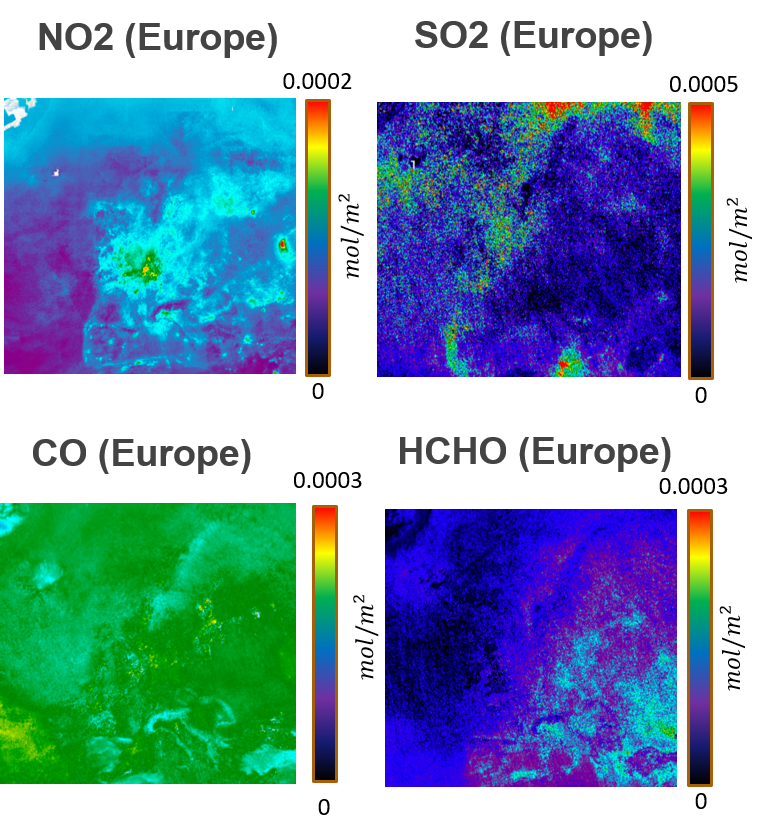}  
  \caption{Sample of Sentinel-5P pollution data}
  \label{fig:s5p_data}
\end{figure}

\begin{figure}[!ht]
    \centering
    \includegraphics[scale=0.6]{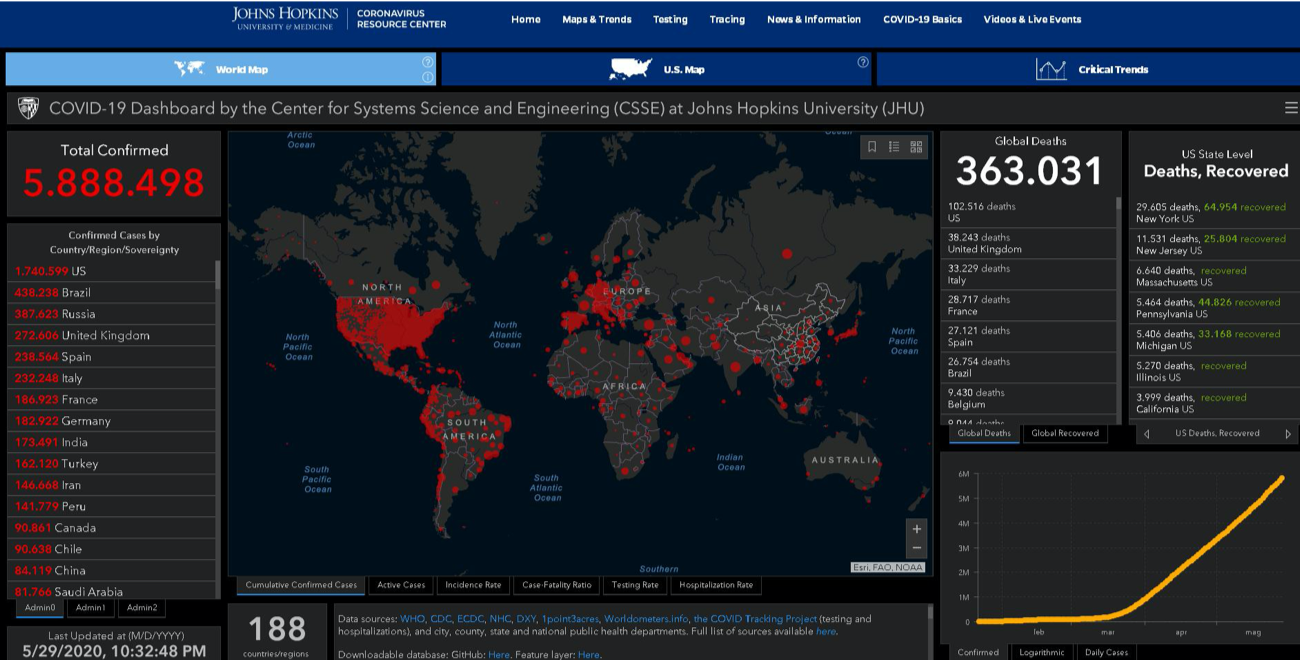}
    \caption{COVID-19 Johns Hopkins Dashboard \cite{coviddashboard}.}
    \label{fig:covid_dashboard}
\end{figure}

The purpose of the measurement of pollutants is twofold. As it will be discussed ahead, some pollutants favor the transmission of the respiratory diseases and therefore by measuring them and feeding the model with their values, a contribution to the final risk level is obtained. In addition, some pollutants give us information on the degree of lockdown. Again, the model is able to trace the goodness of the lockdown by using as input certain types of pollutants, and to capture the inter-relationships among the data.
What is important is to understand that the analysis wants to highlight two different levels of vision: a macro high-level vision (i.e. through the use of satellites, for instance, or data at a wide geographical level, national, etc.), and a micro low-level vision where local understanding of the phenomenon is carried out. 
The proposed tool based on AI and the specific neural network can be used at each level, once trained, and give important insights. An interesting analysis has been done in \cite{Laryetal}, which discusses how ML algorithms can help characterizing airborne particulates for applications related to remote sensing. 
\\
An example of data retrieved both through  ground and satellite platforms  is shown in Fig. \ref{fig:air_quality} and Fig. \ref{fig:s5p_data}, where an air quality map at European level is presented (retrieved through the European Environmental Agency website \cite{europeanmap}) together with some pollutants' levels obtained through the Sentinel-5P satellite, while regarding the COVID-19, the use of the public Johns Hopkins Dashboard (see Fig. \ref{fig:covid_dashboard}) has been proposed, that allows to download the contagion data organized per state, region, cities, and with much more information, such as the daily new cases, daily new deaths and so on, in all the world \cite{coviddashboard}.


\subsection{Characterizing airborne particulates}
About the relationship between pollution and COVID diffusion, there are several schools of thought, and among these, those who consider the causal relationship between pollution and COVID confirmed, therefore believing that it plausible that the coronavirus can be transported in the air and the contagion may happen also at a less than 2 meter distance. Other schools of thought believe that the respiratory system is already put in great difficulty by pollution in some areas, so when COVID arrives it is more likely to affect in a severe way the population and kill it. Moreover, there are two sub-classes of thought:  those who believe that there  can be a chronic factor (i.e. those who are used to being in polluted air are more exposed to serious effects of the virus), and those who instead think that the problem is contextual (i.e. during strong episodes of pollution, the organism is acutely exposed to severe stress, which makes it more vulnerable to the virus).
 
At the moment none of these hypotheses has been incontrovertibly demonstrated, so they could still be all valid, and therefore various factors could be responsible for the spread of the infection.

In any case, we considered very important to analyze the NOx presence in the air, and we are following the idea that particles can carry on the virus, stabilizing it over time, and spreading it in space. For this reason, some work has been done to find a correlation between the virus and the NOx, since there is a virus/dust correlation (the dust may cause the virus transport), and NOx is part of the elements captured and transported by the dust.

\subsection{Particulate Matter (PM)-virus correlation}
More than 240 scientists recently signed a Commentary, appealing to the medical community and highlighting the need of dealing with the airborne transmission of SARS-CoV-2 \cite{morawska2020time}. Many researches have proven that the previously indicated “safe distance” of 6 feet cannot be considered sufficient, since different way of diffusion can occur, in indoor and outdoor environments \cite{setti2020airborne}.
Since the beginning of the COVID-19 pandemic, several studies have been carried out to investigate the reasons of the uneven distribution of infections and fatalities within the different Countries, and positive correlations with air pollution, particularly with suspended fine particles, have been found \cite{frontera2020regional, setti2020potential, magazzino2020relationship, yongjian2020association, sanita2020novel}. Some researchers explain these results considering acute and chronic effects towards the respiratory system, that could make it more susceptible to pathogen infection, while others suggest that different biotic and abiotic factors could be inhaled adhering to the already suspended fine particles \cite{ma2020understanding}.\\
Even though the modes of COVID-19 transmission are still under discussion \cite{al2020sars, farhangrazi2020airborne, tung2020particulate}, the possibility of considering the presence of SARS-CoV-2 RNA on PM10 in outdoor environments has also been suggested \cite{setti2020searching}. Many previous studies already demonstrated that different varieties of microorganisms are present on the suspended particulate \cite{Cao2014inhalable}, and recently early warnings were given about the possibility that they could play a role of carrier for the coronavirus \cite{qu2020imperative}, supported by the finding of SARS-CoV-2 or viral nucleic acid on particulate samples in outdoor environments \cite{liu2020aerodynamic}, \cite{setti2020sars}.\\
According to the cited researches, both PM10 and PM2.5 could be relevant in improving viral infectivity. However, it has been hypothesised that the effect is not linear at all conditions, but that the prolonged high concentration of fine particulate (for instance, more than the daily limit of 50 $\mu$g/m3 of PM10) could trigger a 'boost' effect on the spread of virus \cite{setti2020potential}.\\

\subsection{Wireless Sensors Networks for PM Mapping}
Given the above considerations, and by taking into account that not all the pollutants can be recovered through a satellite analysis, and that however different levels of investigations may become necessary, it is important to discuss the characteristics that the sensor networks must hold in order to guarantee the collection of the pollutant values in a way that is useful for the intended purposes.

In regards to the sensors used at the ground level, the main challenge, and a significant element of innovation, concerns exactly the identification of dust sensors, in particular for PM2.5, of a certain cost, consumption, and footprint for a widespread installation in both fixed and mobile (for example with drones) wireless networks, to investigate   geographical areas of particular interest, while maintaining peculiar characteristics such as selectivity and sensitivity in the measurement.\\
Traditionally, PM is measured by means of the laser scattering technique. Particles are hydro- dynamically focused to flow in a single stream on which a laser beam is focused. The presence and size of each single particle can be determined from the intensity of the scattered light pulse, assuming a spherical shape and average optical properties. The granulometric distribution is thus obtained with reasonable accuracy, and the smallest detectable diameter is diffractionl-imited, in the order of the light wavelength, i.e. hundreds of nm. The laser scattering technique represents the state of the art for particles in the 0.3-10 $\mu$m size range. However, due to their cost and bulkiness these instruments are not suitable for massive deployment in the environment. They are typically installed in a few fixed monitoring stations controlled by local environmental protection agencies.

Given the relevance of air pollution for human health and the need for better spatio-temporal resolution in mapping complex phenomena, such as dust generation, concentration and transport, several efforts were carried out in the last decade, to develop compact and affordable devices for measuring PM in a distributed, pervasive way. Recent implementations of wireless sensors networks at city scale have demonstrated the feasibility of this paradigm \cite{cityscale}, and specific technologies have been leveraged to address several challenges, in particular to preserve the environment and human health, spanning from monitoring air, water \cite{water} and the surrounding environment to control natural disasters or preserving landscape resources \cite{prospects}.

The employment of miniaturized devices has spread out, with different characteristics, mainly grouped into two classes. Micro-machined silicon-based sensors represent the ultimate degree of miniaturization and leverage micro-fabrication capabilities to enhance the detection sensitivity \cite{emerging}. Two main approaches have been proposed for solid-state detection: the use of mechanical resonance in oscillating micro- and nano-weighting scales and high-resolution capacitance measurements of the single-particle impedance on chip \cite{zepto}. Despite very promising preliminary results and potential for nanoparticles detection as well as ubiquitous integration in handheld devices, such as smartphones, thanks to their millimetric size, they appear still far from a commercial maturity. A critical aspect, in addition to clogging, cleaning, lifetime and power dissipation, when shrinking down the sensor size, remains the active fluidics necessary to capture and collect dust particles in the chip.

\begin{figure}[!ht]
    \centering
    \includegraphics[scale=0.5]{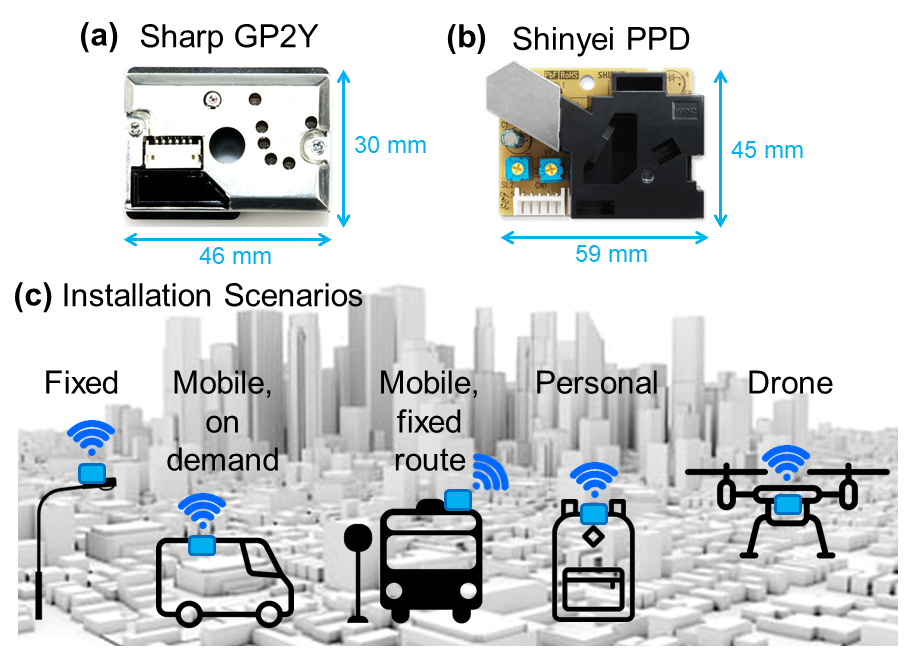}
    \caption{Examples of low-cost credit-card sized PM sensors from Sharp (a) and Shinyei (b) enabling new versatile scenarios (c) in pervasive monitoring of dust concentration, also capable of coping with emergency situations such as the acute phases of pandemics.}
    \label{fig:particolato}
\end{figure}

Another, more consolidated class of PM sensors is that of low-cost optical sensors, named Optical Particle Counter (OPC). They represent an evolution of smoke detectors in which a photodetector measures the amount of scattered light from the dust when  illuminated by a LED. Several sensors of this class are available in the market from companies such as Alphasense, Honeywell, Plantower, Sharp and Shinyei. The latters are very compact (credit-card sized, Fig. \ref{fig:particolato}, (a) and (b)) and consolidated, enabling new versatile scenarios (Fig. \ref{fig:particolato}, (c)) in pervasive monitoring of dust concentration, also capable of coping with emergency situations such as the acute phases of pandemics.

Different works have compared their performance with reference instrumentation, both in the lab \cite{PMLab1, PMLab2} and in the field \cite{fieldtest}, by finding good agreements. In the majority of conditions, they can quantify the concentration of particles with a size range similar to that of laser scattering, with a full scale of about 1000 $\mu g/cm^3$, and a sensitivity of a few tens of $\mu g/cm^3$, well matched with the regulatory limit of 50 $\mu g/cm^3$. Furthermore, the response time is in the orders of seconds, fast enough to capture the dynamics of human and air transport. Thus, they are suitable for the purpose of this project, enabling the deployment of thousands of sensing nodes capable of quickly detecting exceeding of PM10 and PM2.5 concentration limits. Indeed, Artificial Intelligence can leverage the high level of redundancy and overlap in spatial mapping to compensate for the intrinsic limits of this class of devices.

~\newpage


\subsection{Case studies: correlation between NO2 data and COVID-19 related data}

As highlighted before, it is extremely important to change the level of analysis, from a macro to a micro-level, to better understand and monitor the evolution of the phenomenon, and  to get the right input values for the DSS in such a way that the initial diffusion of the virus can be detected and stopped with targeted levels of lockdown. 

\begin{figure}[!ht]
  \centering
  \includegraphics[width=1\linewidth]{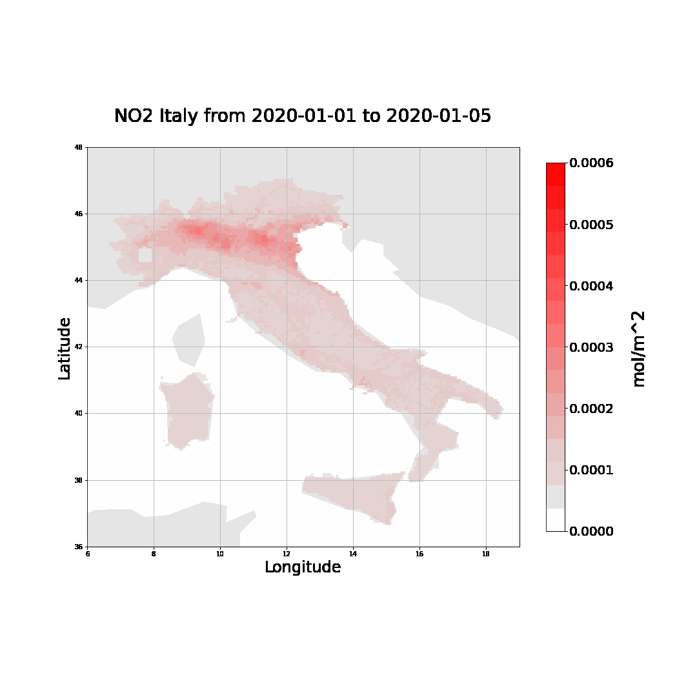}  
  \caption{$NO_2$ concentration in Italy (from Jan. 1st to Jan. 5th 2020)}
  \label{fig:s5p1_1}
\end{figure}

\begin{figure}[!ht]
  \centering
  \includegraphics[width=1\linewidth]{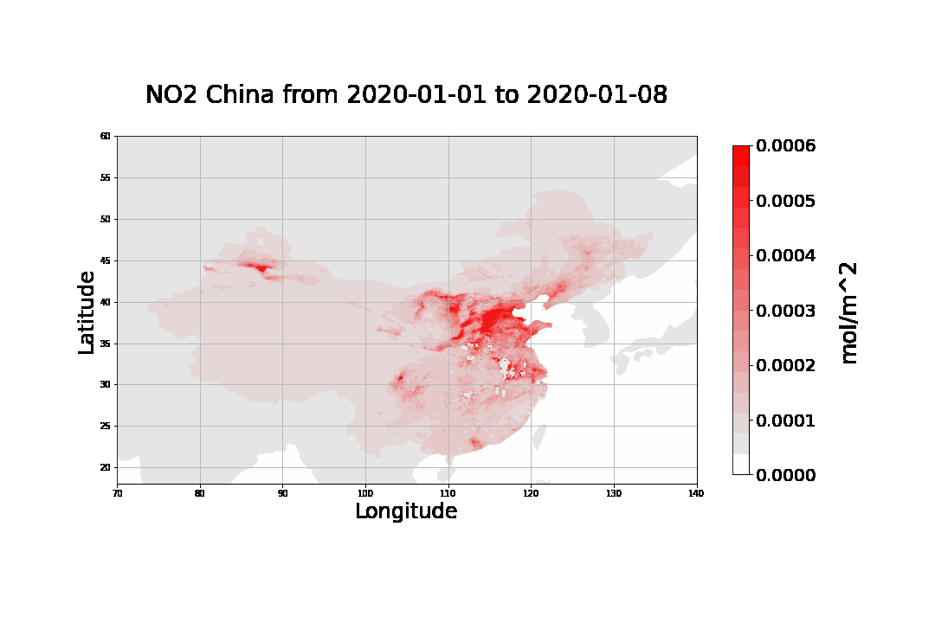}
  \caption{$NO_2$ concentration in China (from Jan. 1st to Jan: 8th 2020)}
  \label{fig:s5p1_2}
\end{figure}

In a first part of the analysis, a study has been carried out to evaluate the correlation between one of the pollutant, the NO2, and the COVID-19.
 The data relating to the  NO2 concentration have been acquired through the Google Earth Engine (GEE), recorded through the use of  the Sentinel-5P satellite,   processed, averaged  and plotted  by  using Python scripts.  In the Fig. \ref{fig:s5p1_1} and  Fig. \ref{fig:s5p1_2} the concentration of NO2 is shown for Italy and China respectively, over two similar periods. It is possible to see that in the worst period of pandemic for both countries, the highest concentration of NO2 laid in the regions where the COVID-19 had its highest values.


For this reason, rather than analyzing the data relating to entire nations, we have chosen to focus on the regions of Italy and China where the spread of the virus has been most rapid and emblematic (Lombardy and the Wuhan area).
Sentinel-5P covers wide areas, therefore, by starting from the national concentrations of nitrogen dioxide,  a subset to obtain the concentrations related to the chosen case studies has been created. Some statistics on the subset have been calculated, including the maximum, the minimum, the standard deviation, for instance, and maximum values of the pollutant have been collected in the following Table 1 and Table 2 for further discussion. In the two tables, to manage the absence of values in Sentinel-5P data, the average values of concentrations for the NO2 over five days have been calculated, by starting from January 1rst until June 30th, 2020.

The correlation between the peaks of the average NO2 concentration and the number of new COVID-19 positives has been analyzed. To match the satellite and epidemiological data, it was necessary to make an average over 5 days for the number of new infections. A scatter plot was constructed using these data, positioning the number of infected people on the abscissa axis and the $NO_2$ concentration on  the ordinates.

\begin{figure}[!ht]
	\centering
	\includegraphics[scale=0.7] {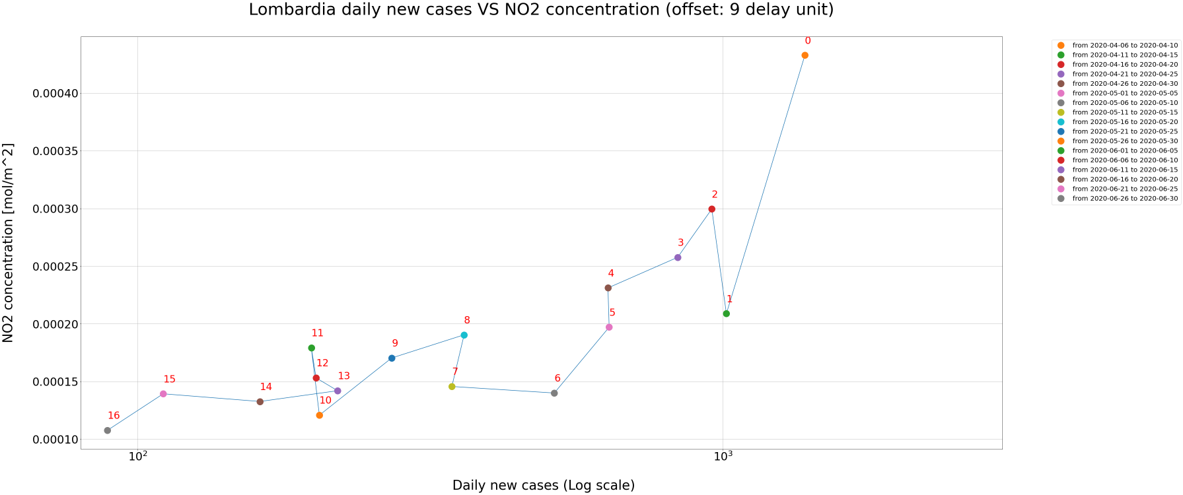}
	\caption{New daily cases VS average concentration of $ NO_2 $ in Lombardy}
	\label{lombardiadelaymax1}
\end{figure}

\begin{figure}[!ht]
	\centering
	\includegraphics[scale=0.8] {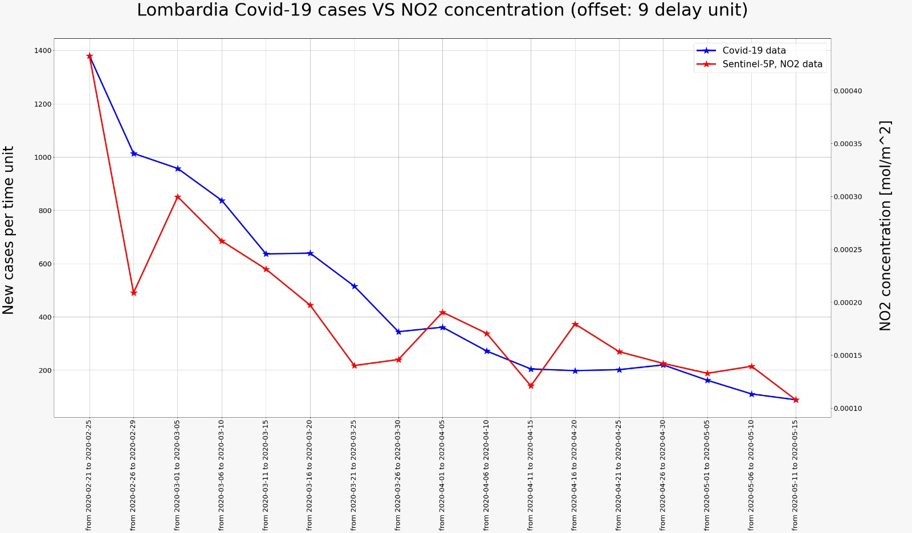}
	\caption{New daily cases VS average concentration of $ NO_2 $ in Lombardy}
	\label{lombardiadelaymax2}
\end{figure}

\begin{figure}[!ht]
	\centering
	\includegraphics[scale=0.7] {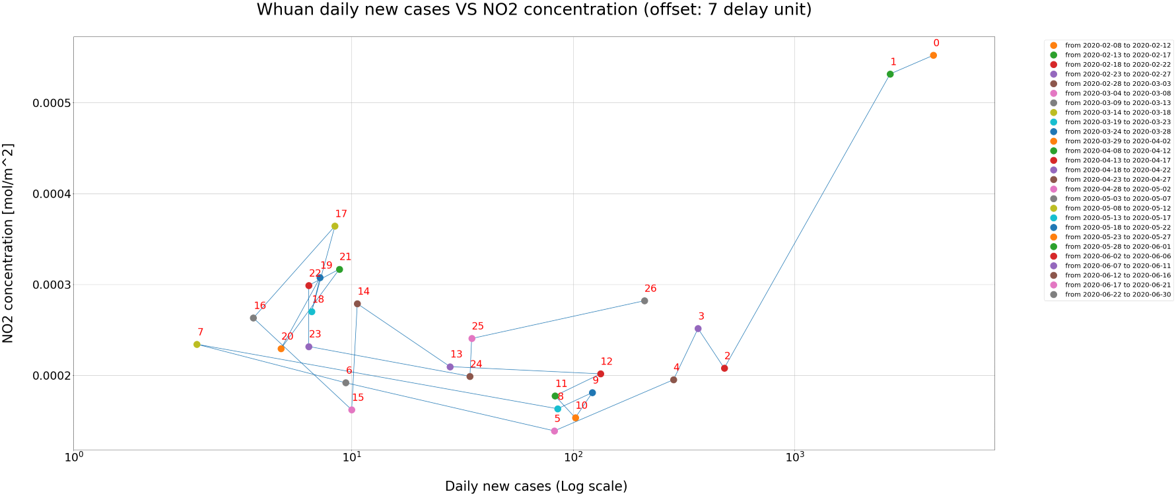}
	\caption{New daily cases VS average concentration of $ NO_2 $ in Wuhan.}
	\label{wuhandelaymax1}
\end{figure}

\begin{figure}[!ht]
	\centering
	\includegraphics[scale=0.8] {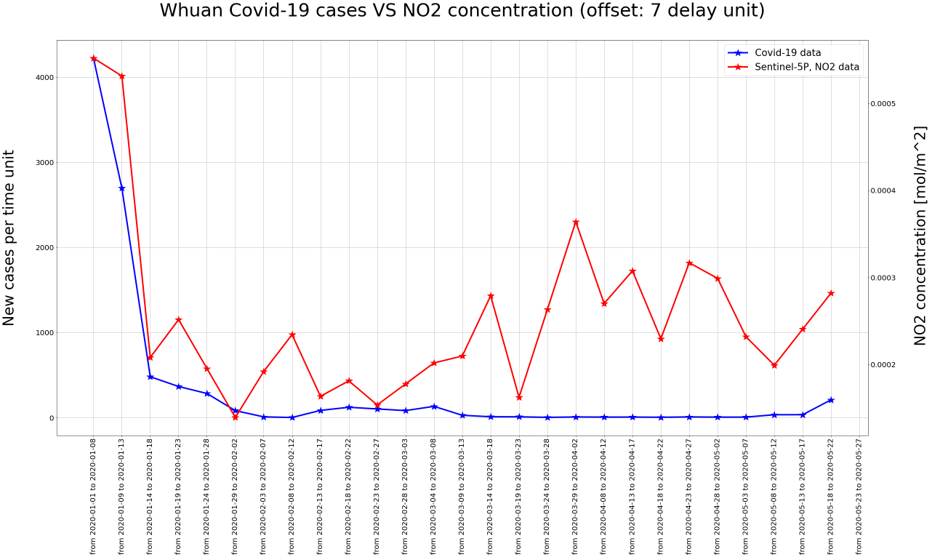}
	\caption{New daily cases VS average concentration of $ NO_2 $ in Wuhan}
	\label{wuhandelaymax2}
\end{figure}

 In the Fig. \ref{lombardiadelaymax1} and Fig. \ref{wuhandelaymax1}, the scatter plots present a time sequence of data, where the time is defined by the number near the dot. This representation helps to graphically catch a correlation between  NO2 and COVID-19 and also to identify emblematic cases like the Wuhan ones. In fact, in this latter case the trend, at a certain moment, starts oscillating. This phenomenon  can be better explained by analyzing the Fig. \ref{lombardiadelaymax2} and Fig. \ref{wuhandelaymax2}, where it is evident that for the Lombardy region  both the NO2 and COVID-19 have a decreasing trend in time (from a  high number of infected and a  high concentration of pollutant to a low number of infected and a low concentration of pollutant), while for the Wuhan region the situation is quite different and unexpected, since after a certain period the NO2 starts increasing again while the new infections are stable and close to a very low value. 

To take into account the delay between the instant when infection occurs and the actual evidence of the transmitted contagion a delay analysis has been carried out: the COVID-19 data were moved forward in time, obtaining scatter plots with different delays. The correlation was calculated using the Pearson correlation coefficient (PCC), a statistic that measures the linear correlation between two variables X and Y, with a value between -1 and +1. A value of +1 indicates total positive linear correlation, 0 indicates no linear correlation, and -1 indicates total negative linear correlation.

\begin{table}[!ht]
\centering
 \begin{tabular}{|c|c|c|} 
 \hline
 Delay unit & PCC Lombardia & PCC Wuhan \\ 
 \hline\hline
    0  &  0.0770 & -0.2474\\
    1  &  0.2773 & -0.2514\\
    2  &  0.4983 & -0.2496\\
    3  &  0.6629 & -0.1775\\
    4  &  0.7180 &  0.0271\\
    5  &  0.7918 &  0.1934\\
    6  &  0.8774 &  0.3842\\
    7  &  0.8092 &  \textbf{0.4857}\\
    8  &  0.7532 &  0.3905\\
    9  &  \textbf{0.8969} &  0.2969\\
    10 &  0.8334 &  0.2905\\
    11 &  0.8403 &  0.1813\\
    12 &  0.8736 & -0.0652\\
    13 &  0.7830 & -0.2000\\
    14 &  0.8302 &  0.0291\\
    15 &  0.8854 &  0.1762\\
 \hline
 \end{tabular}
 \caption{Pearson Correlation Coefficient for Lombardia and Wuhan}
 \label{correlazioneTable2}
\end{table}

From Table \ref{correlazioneTable2}  it can be seen that the maximum correlation (positive) was recorded with a delay of 9 units (Fig. \ref{lombardiadelaymax1} and Fig. \ref{lombardiadelaymax2}) in the case of Lombardy region.

Similarly, from Table \ref{correlazioneTable2}  it can be seen that the maximum correlation (positive) was recorded with a delay of 7 delay units (Fig. \ref{wuhandelaymax1} and Fig. \ref{wuhandelaymax2}) in the case of Wuhan.

Several considerations can be made.

First of all, if the delay units are  converted back  to number of days, the maximum value of the correlation between NO2 and COVID-19 is found after 35 days for Wuhan region, and for 45 days for Lombardy region. Somehow, it seems that restrictions in Wuhan brought to better results in terms of COVID-19 reduction than in Italy. Moreover, it is clear that the \textit{a-posteriori}  measures of lockdown cannot be an efficient mode of intervention, because before the effects on infection reduction become significant  more than one month, one month and half must pass. Lastly,  if the graphs related to Wuhan are analyzed (as underline before) an anomaly appears evident. In Fig. \ref{wuhandelaymax2}, when the NO2 values begin to increase (since the lockdown has been removed), the number of new infected remains very low, and this  trend raises some doubts about the veracity of the data on the number of infected communicated by Wuhan in this second phase. Clearly a further analysis   might give additional insights.

\section{Conclusions}

In this paper we have presented the cross-disciplinary AIRSENSE-TO-ACT project, aiming at the creation of a Decision Support System for the timely and effective activation of targeted  countermeasures during virus pandemics, based on a model merging data from very heterogeneous sources, including ground wireless networks of low-cost dust monitors, and satellite data, spanning from meteorological and pollution data to crowd sensing. The correlation between virus diffusion and concentration of NO2 has been analyzed, and the further analysis of the correlation between virus diffusion and PM in the air would be the main focus of future investigations. Moreover, work is in progress   on the integration of mobility data \cite{mobility} inside the model. 

The "second wave" currently spreading in Europe demonstrates the urgent need for such a tool as proposed in this paper, in order to limit the economical damage of generalized lockdowns and restrictions.

\section{Acknowledgments}
Authors want to thank Maria Pia Del Rosso and Chiara Zarro for their useful insights and support.

\authorcontributions{All the authors contributed equally to this work}



\conflictsofinterest{The authors declare no conflict of interest.}

\reftitle{References}
\externalbibliography{yes}
\bibliography{main}

\end{document}